\documentclass[twocolumn,aps,pra]{revtex4}
\usepackage{amsmath,amssymb,mathrsfs,bm}
\usepackage{graphicx,dcolumn,times}
\usepackage{xcolor}

\begin{document}
\title{$\delta$-Quench measurement of quantum wavefunction}
\author{Shanchao Zhang$^1$\footnotemark[2], Yiru Zhou$^1$\footnotemark[2],  Yefeng Mei$^2$\footnote[2]{These authors contributed equally.}, Kaiyu Liao$^1$, Yongli Wen$^3$, Jianfeng Li$^1$, Xin-Ding Zhang$^1$, Shengwang Du$^{2,1}$\footnotemark[1], Hui Yan$^1$\footnotemark[1] and Shi-Liang Zhu$^{3,1}$\footnote[1]
{email: dusw@ust.hk; yanhui@scnu.edu.cn; slzhu@nju.edu.cn}}
\affiliation{$^1$Guangdong Provincial Key Laboratory of Quantum Engineering and Quantum Materials, GPETR Center for Quantum Precision Measurement and SPTE, South China Normal University, Guangzhou 510006, China\\
$^2$Department of Physics \& William Mong Institute of Nano Science and Technology, The Hong Kong University of Science and Technology, Clear Water Bay, Kowloon, Hong Kong S.A.R., China\\
$^3$National Laboratory of Solid State Microstructures, School of Physics, Nanjing University, Nanjing 210093, China}
\begin{abstract}
Measurement of quantum state wavefunction not only acts as a fundamental part in quantum physics but also plays an important role in developing practical quantum technologies. Conventional quantum state tomography has been widely used to estimate quantum wavefunctions, which, however, requires huge measurement resources exponentially growing with the dimension of state. The recent weak-value-based quantum measurement circumvents this resources issue but relies on an extra pointer space. Here, we propose and demonstrate a simple and direct measurement strategy based on $\delta$-quench probe: By quenching its complex probability amplitude one by one ($\delta$-quench) in the given bases, we can directly obtain the quantum wavefunction by projecting the quenched state onto a post-selection state. As compared to the conventional approaches, it needs only one projection basis. We confirm its power by experimentally measuring photonic complex temporal wavefunctions. This new method is versatile and can find applications in quantum information science and engineering.
\end{abstract}\maketitle

\section{Introduction}
Unlike the measurement of a classical quantity that relies on deterministic and accurate instrument readings, the measurement outcomes of a quantum state are probabilistic \cite{vonNeumann1955} and follow the Heisenberg uncertainty principle. As the projections on one measurable bases give only real numbers, to determine the complex quantum wavefunction of a quantum state requires additional resources and is often challenging.

Quantum state tomography (QST) is a conventional way to estimate a quantum wavefunction by fitting data from plenty projection measurement onto at least two sets of bases with multi-parameters estimation algorithms\cite{James2001,Ariano2003,Sosa2017,Smithey1993, Breitenbach1997, Resch2005}. However, as the dimension of the Hilbert space increases, QST demands exponential-growth resources and eventually becomes exhaustive and infeasible. Recently, the efficient weak-value-based quantum measurement (WQM) \cite{Aharonov1988,Dressel2014,Lundeen2011,Vallone2016,Lundeen2012,Salvail2013,Thekkadath2016,Calderaro2018,Malik2014,Kocsis2011,Tobias2018,Hacohen2016} is developed to directly read the quantum wavefunctions. This strategy relies on an ancillary pointer state space and its coupling with the unknown quantum state, which makes the strategy sometime invalid\cite{Haapasalo2011,Maccone2014} due to, for example, the absence of a suitable pointer.

\begin{figure}[ptb]\begin{center}\includegraphics[width=8.5cm]{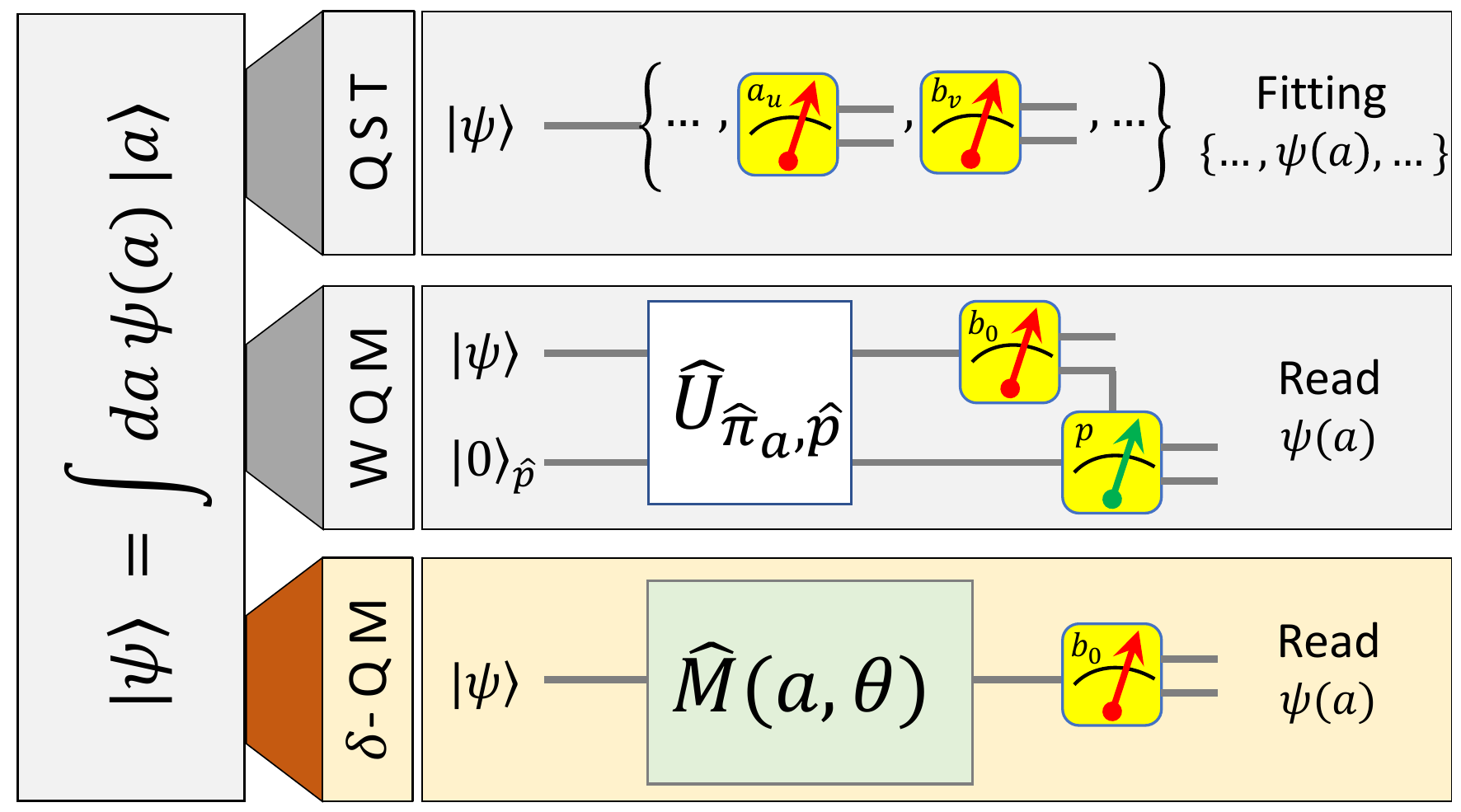}
\caption{\label{fig1} \textbf{Schematic comparison of quantum measurement strategies.}
To determine the wavefunction in the bases $\{|a\rangle\}$, conventional quantum state tomography (QST) needs a collection of projection measurement in $\{|a\rangle\}$ and its mutually unbiased bases(MUB) $\{..,\{|b_v\rangle\},...\}$ followed by multi-parameters fitting process. Weak-value-based quantum measurement (WQM) couples ($\hat{U}_{\hat{\pi}_a,\hat{p}}$ with $\hat{\pi}_a=|a\rangle\langle a| $) the quantum state with a pointer state ($|0\rangle_{\hat{p}}$) then directly read $\psi(a)$ from the pointer state measurement conditioned on a post-selection projection. Our proposed $\delta$-quench measurement( $\delta$-QM) directly read $\psi(a)$ sequentially by firstly quenching the quantum state($\hat{M}(a,\theta)$) and then analysing the response in projection measurement of the quenched state onto only one fixed post-selection state($|b_0\rangle$) belongs to a MUB of bases $\{|a\rangle\}$.}
\end{center}\end{figure}

\begin{figure*}[ptb]\begin{center}\includegraphics[width=18cm]{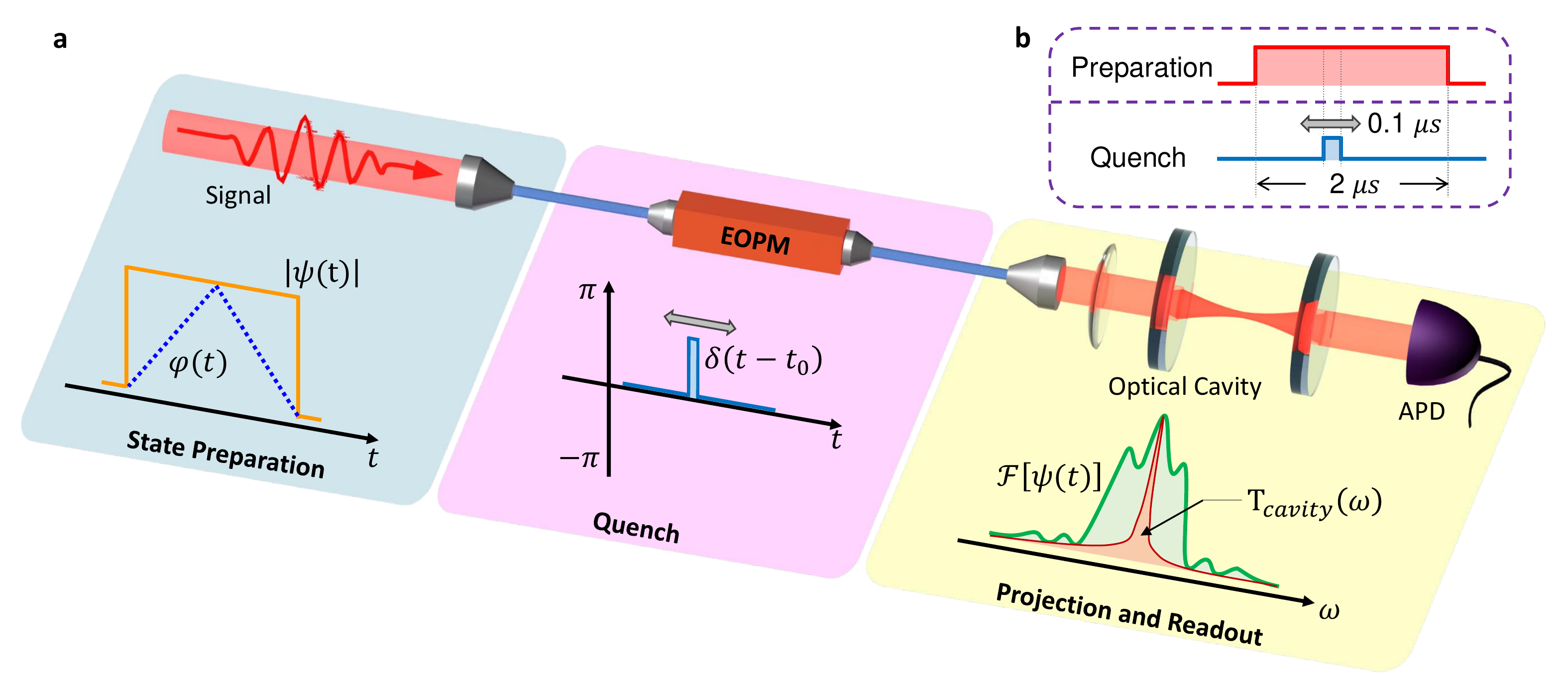}
\caption{\label{fig2} \textbf{Experimental setup for the $\delta$-quench measurement of the temporal wavefunction.}
\textbf{a.} Signal photons are produced by attenuating a laser beam with central wavelength $\lambda=$780 nm (carrier frequency $\omega_0=2\pi c/\lambda$ with $c$ as the light speed) and line-width of around 50 $\rm kHz$. The average power of photon stream is about 10 $\rm \mu W$. The amplitude of complex envelope $|\psi(t)|$ is prepared with an acousto-optic modulator (Brimrose, TEF-110-30-780). A fiber based fast electro-optical phase modulator (EOPM, EO-Space, PM-0K5-20-PFU-PFU-795-UL) with modulation bandwidth of 20GHz is used to both prepare the phase envelope $\varphi(t)$ and realize the $\delta$-like phase quench ($e^{i\theta}$). The post-selection projection measurement is conducted with a high finesse optical cavity (finesse $F$=20000, line-width of $72 \rm kHz$, Customized by Stable Laser System), whose resonance frequency is tuned to be $\omega_0$. The transmitted photons are eventually detected by an avalanche photon detector (APD, Thorlabs, APD120A/M) and then recorded by an oscilloscope (Tektronics, DPO4104B). 
\textbf{b.} The temporal length of photonic quantum wavefunction is 2$\mu s$ and the measurement time-bin width is 0.1$\mu s$.}
\end{center}\end{figure*}
In this work, we propose and demonstrate the $\delta$-quench measurement ($\delta$-QM) method, a new type of versatile strategy for quantum wavefunction measurement.  For an unknown quantum state, we quench it by varying one of its complex probability amplitude in a measurement Hilbert space and then project the quenched state onto a post-selection state that is non-orthogonal to all the bases in the measurement space. The real and imaginary components of its quantum wavefunction can be directly obtained from the sequentially measured quench dependent responses. As compared to QST and WQM, the $\delta$-QM requires projection measurement on only one fixed post-selection state and does not need parameter estimation algorithms, as summarized in Fig.~\ref{fig1}. As an example, we use $\delta$-QM to experimentally measure photonic complex temporal wavefunctions, which has, up to now, not yet been directly measured\cite{Wasilewski2007, Beduini2014, PChen2015,YangPRA2018,Davis2018, Ansari2018}. Our experimental results verify that this method is robust and efficient with limited measurement resources.

\bigskip
\begin{figure*}\begin{center}
\includegraphics[width=18cm]{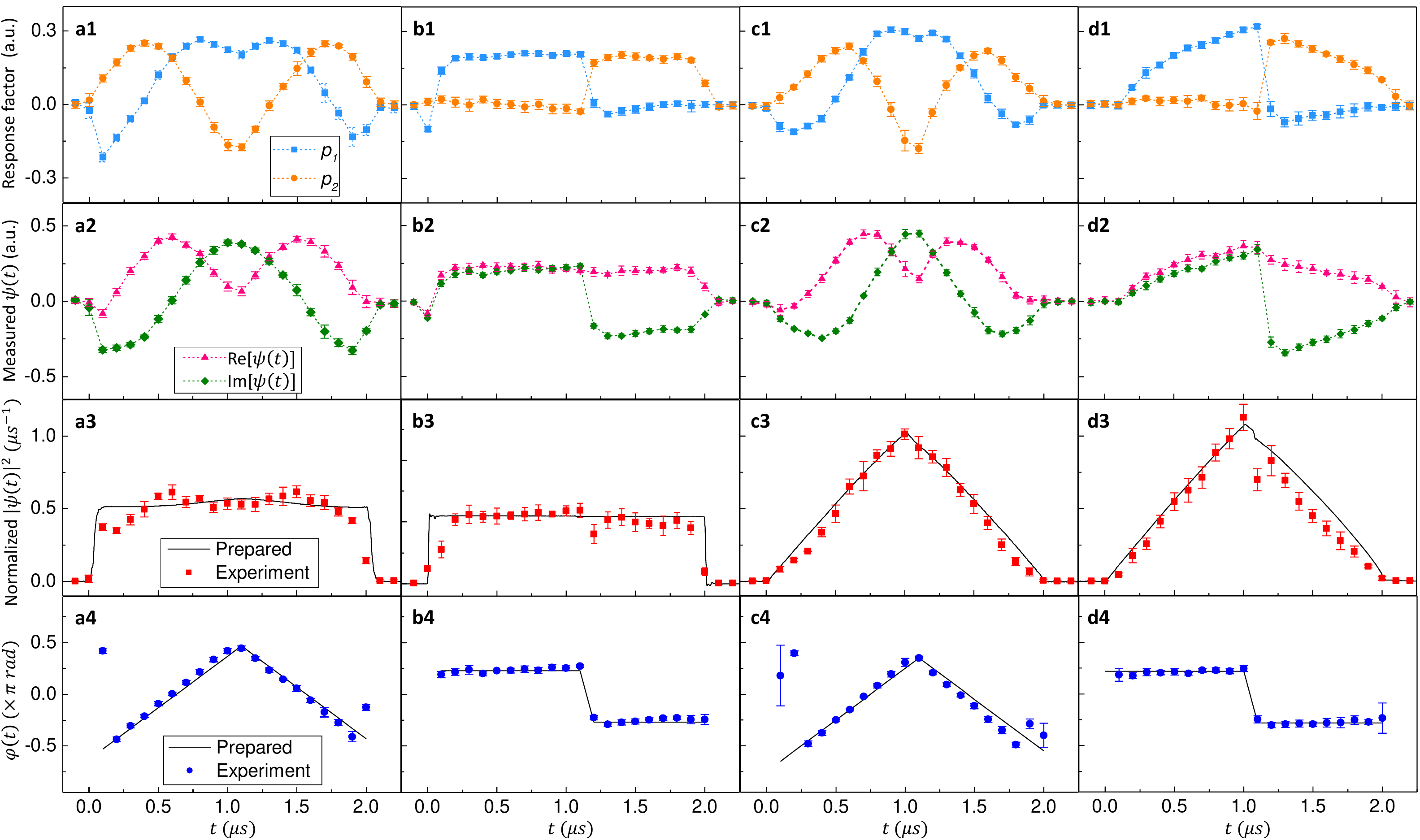}
\caption{\label{fig3}\textbf{The measured response factors and temporal wavefunctions.}
\textbf{a1-d1,} Experimentally measured response factor $p_{1,2}$ for the prepared temporal wavefunction with different envelopes. 
\textbf{a2-d2,} Directly calculated real and imaginary components of temporal wavefunction $\text{Re}[\psi(t)]$ (pink triangles) and $\text{Im}[\psi(t)]$ (green diamonds). \textbf{a3-d3,} Normalized intensity envelopes of temporal wavefunction $|\psi (t)|^2$. \textbf{a4-d4,}  Phase envelops obtained by equation $\varphi(t)=\arctan (\text{Im}[\psi(t)]/\text{Re}[\psi(t)])$. Dashed lines in \textbf{a1-d1} and \textbf{a2-d2} are just eye-guide lines connect neighbouring data points. Solid lines in \textbf{a3-d3} are the normalized intensity profiles of quantum state measured before being quenched. Solid lines in \textbf{a4-d4} are the theoretical phase envelopes derived from the preparation equipments. Error bars denote the statistical variance of 1 s.d.}
\end{center}\end{figure*}

\section{Results}
\subsection{Theoretical model}
A given pure quantum state $|\psi\rangle$ can be measured in a given Hilbert space spanned by a complete orthogonal bases $\{|a\rangle\}$, in which we aim to determine the complex quantum wavefunction $\psi(a)=\langle a|\psi\rangle$. The measurement bases $\{|a\rangle\}$ are the complete eigenstate set of an observable operator $\hat{A}$, which satisfies $\hat{A}|a\rangle=a|a\rangle$. Usually projection measurements on $\{|a\rangle\}$ give only the probability distribution $\{|\psi(a)|^2\}$. We here propose a strategy for measuring the complex $\psi(a)$ sequentially. Without loss of universality, we first discuss our model by assuming the Hilbert space to be continuous.

We quench the quantum state $|\psi\rangle$ by the following operator:
\begin{eqnarray}\label{eq:quenchOperator}
\hat{M}(a,\theta)&=&\hat{I}+\int da' \delta(a'-a)(e^{i\theta}-1) |a'\rangle\langle a'|,
\end{eqnarray}
where $\hat{I}$ is the identity operator, $\theta$ is a real phase. $\delta(a-a')$ is the Dirac delta function and hence we name this method $\delta$-QM. Then we project the quenched state onto a post-selection state $|b_0\rangle$, which is chosen from the bases that are mutually unbiased with $\{|a\rangle\}$ \cite{Bandyopadhyay2002,Weigert2008,Lundeen2012,Thekkadath2016}.
The probability of post-selection projection $\text{Pr}(a,\theta)=|\langle b_0|\hat{M}(a,\theta)|\psi\rangle|^2$ is given as below:
\begin{eqnarray}\label{EqPrC}
\text{Pr}(a,\theta)&=&|\langle b_0|\psi\rangle|^2\big{|}1+\frac{(e^{i\theta}-1)\langle b_0|a\rangle}{ \langle b_0|\psi\rangle}\psi(a)\big{|}^2. \\ \nonumber
\end{eqnarray}
It is critical to note that $\langle b_0|\psi\rangle$ does not depend on $\theta$ or $a$ and $\langle b_0|a\rangle$ that is known in advance for specific $|b_0\rangle$ is independent of $|\psi\rangle$. Therefore, the unknown complex $\psi(a)$ can be yielded by just choosing different non-zero quench phases $\{\theta_1,\theta_2\}$ and detecting the corresponding response factor of projection outcomes $p(a,\theta)=1-\text{Pr}(a,\theta)/P_0$ with $P_0=|\langle b_0|\psi\rangle|^2$ as the measurement outcome without quench.

As an example, in the following we describe how to measure temporal complex wavefunctions of photons $\psi(t)e^{-i\omega_0 t}$, where $\omega_0$ is the carrier optical angular frequency and $\psi(t)$ is complex envelop. Here, we take $|b_0\rangle=|\omega_0\rangle$ for the sake of simplicity and this experimental post-selection projection measurement is realized by a fixed optical frequency filter. To measure $\psi(t)$ at a time instant $t_0$, the phase quenches $\theta=\{\pi/2,-\pi/2\}$ are applied at $t_0$, and the corresponding response factor $p(t_0,\theta)=1-\text{Pr}(t_0,\theta)/P_0$ are denoted as $\{p_1, p_2\}$, respectively, where $P_0$ is the projection outcome without quench and thus independent of $t_0$. We can derive the real and imaginary parts of the wavefunction envelop as below by omitting a normalization factor:
\begin{eqnarray}\begin{array}{ll}\label{EqTdDM}
&\text{Re}[\psi(t_0)]=2-\sqrt{4(1-p_1-p_2)-(p_1-p_2)^2},\\
&\text{Im}[\psi(t_0)]=p_1-p_2.
\end{array}\end{eqnarray}
By stepping $t_0$ sequentially, we are able to obtain the whole wavefunction following the same procedure.

From the above detailed theoretical model description, it is obvious that this $\delta$-QM method can directly obtain the quantum wavefunction and needs only one post-selection projection state. This method can be applied to the measurements of various quantum systems, such as spatial wavefunctions, photonic polarization states, etc. The extension of this method to the discrete Hilbert space is also straightforward (See methods).

\subsection{Experimental results}
The detailed experimental setup for measuring photonic temporal wavefunction using $\delta$-QM method is sketched in Fig.~\ref{fig2}. Photons attenuated from a laser beam are firstly prepared with a test complex temporal wavefunction $\psi(t)e^{-i\omega_0 t}$ and then phase quenched by an electro-optic phase modulator (EOPM) driven by a $\delta$-like impulse in time domain. A high-finesse optical cavity is used as a fixed optical frequency filter to make the post-selection projection measurement. The projection measurement outcomes $\text{Pr}(t,\theta)$ are detected using a fast avalanche photon detector (APD) that is placed after the optical cavity. By changing the depth ($\theta=\{0,\pm\pi/2\}$) of the $\delta$-quench and also stepping the relative time instant of quench ($t_0$), the real and imaginary components of the temporal wavefunction can be obtained. 

We demonstrate the $\delta$-QM method by measuring four different test temporal wavefunctions, as plotted in Fig.~\ref{fig3}. We first sequentially measure the response factors $p_1$ and $p_2$ at each quench instant $t_0$, which are shown in Fig.\ref{fig3}(a1-d1). Following Eqs.(\ref{EqTdDM}), we may directly calculate the real and imaginary parts of the temporal wavefunction $\psi(t)$, as presented in Fig.\ref{fig3}(a2-d2). Straightforwardly, the normalized intensity envelopes $|\psi(t)|^2$ and the phase envelopes $\varphi(t)$ can be directly obtained, as plotted in Fig.\ref{fig3}(a3-d3) and Fig.\ref{fig3}(a4-d4), respectively. It is obvious that measurement results (markers) agree well with the prepared test (solid curves) wavefunctions. Here, the prepared intensity waveforms are measured with APD and the prepared phase envelopse are derived from the electric waveforms that drive EOPM.

\section{Discussion}
Although Eqs.(\ref{EqTdDM}) theoretically work for arbitrary quench depths $\{\theta_1,\theta_2\}$, in reality the quench response factors $p_1$ and $p_2$ are required to be larger than the normalized background fluctuation $\Delta p_0=\Delta P_0/P_0$, where $\Delta P_0$ is the standard deviation of $P_0$. In our experiment $\Delta p_0$ is around $0.002$, which is contributed by the electronic noises from APD and oscilloscopes, the transmission drift of optical cavity and power fluctuation of laser. Therefore, we here also studied the dependence of measurement fidelity(see Methods) on the quench depth $\theta$, which characterizes the performance of this method. 

\begin{figure}\begin{center}\includegraphics[width=8.5cm]{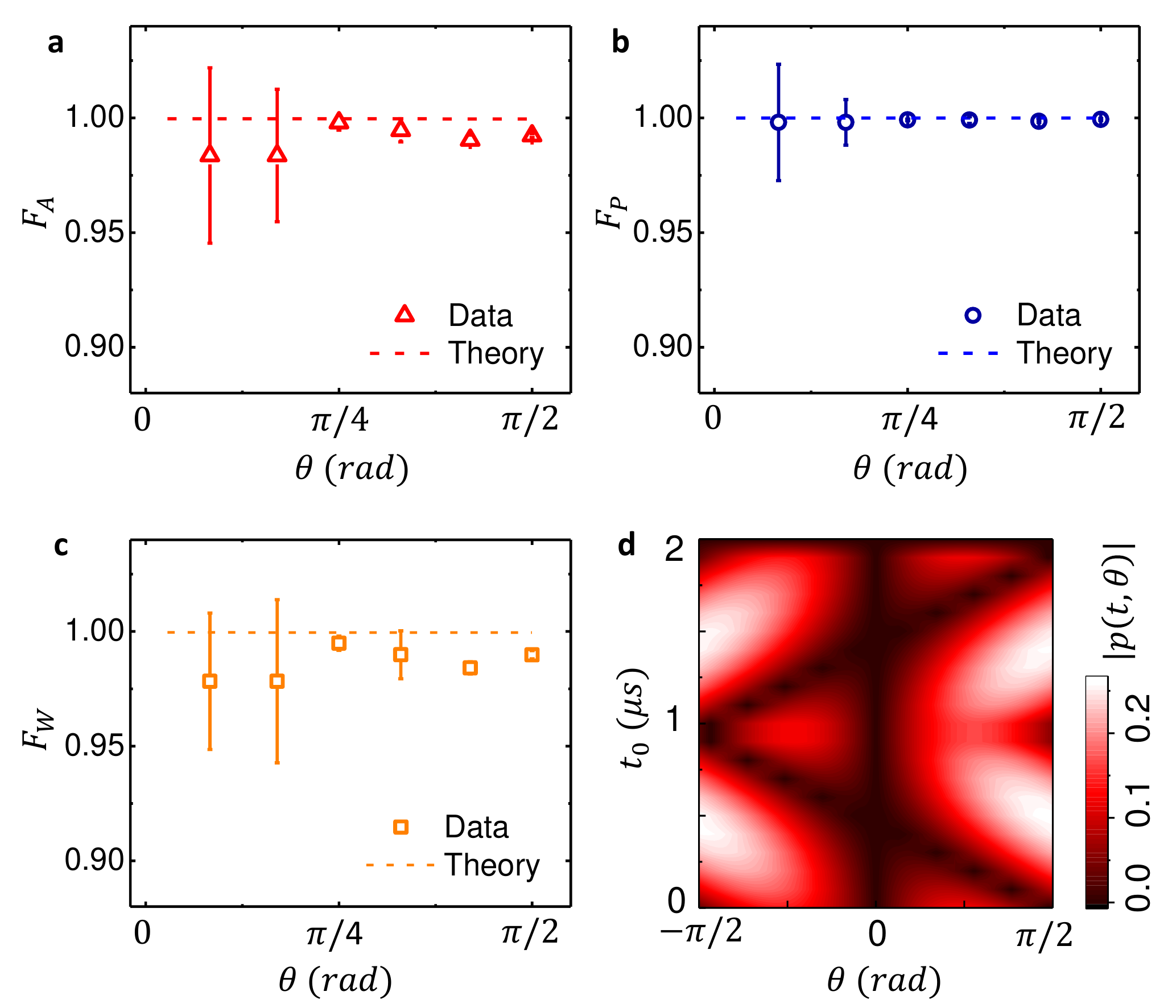}
\caption{\label{fig4}\textbf{The performance of the $\delta$-quench measurements.}
\textbf{a-c,} Fidelities dependence on quench depth of the amplitude, phase and the overall wavefunction, respectively. \textbf{d,} The magnitude of response factor as a function of the quench instant $t_0$ and quench depth $\theta$. The theoretical curves in \textbf{a-c} are the numerically calculated results. Error bars of the experimental data denote the statistical variance of 1 s.d.}
\end{center}\end{figure}

The fidelities of the measured envelopes of amplitude $F_A$, phase $F_P$ and the overall wavefunction $F_W$ as function of the quench depth are plotted in Fig.\ref{fig4}(a-c). At a large quench depth ($\theta\geq\pi/4$), both the fidelities of measured amplitude and phase are maintained as high as nearly unity. The overall fidelity behaves similarly, as shown in Fig.~\ref{fig4}c. When the quench depth is smaller, all the above measurement fidelities decrease and show bigger fluctuation, which can be attributed to the lower signal-to-noise ratio $p_{1,2}/\Delta p_0$.

The detailed relation of response factor and quench depth that determines the measurement fidelity is further investigated numerically. The response factor $p(t,\theta)$ as a function of quench instant $t_0$ and quench depth $\theta$ is calculated with a time-bin width of 0.1$\mu s$ for the temporal wavefunction shown in Fig.~\ref{fig3}(a1-a4). We plot the absolute value of calculated response factors $|p(t,\theta)|$ in Fig.~\ref{fig4}(d). It is obvious that the response factor shows a trend of being positively proportional to the quench depth and the maximum response factor can beyond $0.2$ when the quench depth reaches $\frac{\pi}{2}$, which is 100 times large than the minimum resolvable response factor ($\Delta p_0$=0.002). Considering the results in fig.~\ref{fig4}(a-c), a large range of quench depth can be chosen to keep a high measurement fidelity, which implies the robustness of this strategy.

In summary, we have proposed and demonstrated the $\delta$-QM of quantum wavefunctions as a new versatile measurement method. By sequentially quenching the phase of the bases, we have shown that the quantum wavefunction can be obtained using only one fixed post-selection state. We have applied this method to measure various photonic temporal wavefunctions and achieved the measurement fidelity of around 99$\%$. This $\delta$-quench protocol can also be used to measure wavefunctions in other Hilbert spaces, such as the spatial, polarization and orbital angular momentum degree of freedom. In addition, in spite of that the quench operation is assumed to be unitary above, it is, nevertheless, worth to notice that this strategy can also work with non-unitary quench operation, such as controllable loss, which is currently beyond a conventional quantum measurement explanation and would potentially inspire interesting discussions.

\section{Methods}
\renewcommand{\theequation}{S\arabic{equation}}
\setcounter{equation}{0}
\renewcommand{\thefigure}{S\arabic{figure}}
\setcounter{figure}{0}

\subsection{$\delta$-QM in discrete Hilbert space}
A given pure quantum state can be denoted as $|\psi\rangle=\sum_u \psi_u |a_u\rangle$ in a discrete Hilbert space with complete orthonormal bases $\{|a_u\rangle\}$. It is obvious that $\sum_u |\psi_u|^2=1$. Here the discrete measurable bases $\{|a_u\rangle\}$ contains all the eigenstates of an observable operator $\hat{A}$, which satisfies $\hat{A}|a_u\rangle=a_u|a_u\rangle$. To determine a particular $\psi_n$, we quench the quantum state $|\psi\rangle$ by the operator $\hat{M_n}(\theta)=\hat{I}+\sum_{u} \delta_{n,u}(e^{i\theta}-1)|a_u\rangle\langle a_u|$, with $\hat{I}$ as the identity operator, $\theta$ as a real phase and $\delta_{n,u}$ being the Kronecker delta function. The success probability of projecting quenched state onto a fixed post-selection state $|b_0\rangle$ is $\text{Pr}_n(\theta)=|\langle b_0|\hat{M_n}(\theta)|\psi\rangle|^2$ as below:
\begin{eqnarray}\label{EqPrD}
\text{Pr}_n(\theta)&=&|\langle b_0|\psi\rangle|^2\big{|}1+\frac{(e^{i\theta}-1)\langle b_0|a_n\rangle}{ \langle b_0|\psi\rangle}\psi_n\big{|}^2, \\ \nonumber
\end{eqnarray}
where $|b_0\rangle$ can be chosen from the mutually unbiased bases of $\{|a_u\rangle\}$ and thus make $\langle b_0|a_n\rangle$ an aforehand known non-zero factor before measurement.  Furthermore, $\langle b_0|\psi\rangle$ is independent of $n$ and $\theta$ and can also be chosen non-zero. Therefore, the unknown complex $\psi_n$ can be yielded by detecting the corresponding response factor of projection outcomes $p_n(\theta)=1-\text{Pr}_n(\theta)/P_0$ in the following way. Here, $P_0=|\langle b_0|\psi\rangle|^2$ is the measurement outcome without quench.  Especially, by choosing $|b_0\rangle=B_0\Sigma_u|a_u\rangle$ with $B_0$ as a normalization factor, $\langle b_0|a_u\rangle$ becomes a real constant. With $\theta=\{0, \pi/2, -\pi/2\}$, we have below detailed response factors:
 \begin{eqnarray}\label{EqDSupp}\begin{array}{ll}
P_0\rightarrow\text{Pr}_n(0)&=|\langle b_0|\psi\rangle|^2,\\
p_1\rightarrow1-\text{Pr}_n(\pi/2)/P_0&=1- |1+(i-1)\psi_{n}B_0/\langle b_0|\psi\rangle|^2,\\
p_2\rightarrow1-\text{Pr}_n(-\pi/2)/P_0&=1-|1-(i+1)\psi_{n}B_0/\langle b_0|\psi\rangle|^2.\\
\end{array}\end{eqnarray}
With the normalization condition of $|\psi\rangle$, real and imaginary components of $\psi_{n}$ can be easily obtained as below by omitting the constant normalization factor $\sqrt{P_0}/4B_0$ and an global trivial reference phase factor $\langle b_0|\psi\rangle/|\langle b_0|\psi\rangle|$:
\begin{eqnarray}\label{EqDSupp2}\begin{array}{ll}
\text{Re}[\psi_{n}]&=2-\sqrt{4(1-p_1-p_2)-(p_{1}-p_2)^2},\\
\text{Im}[\psi_{n}]&=p_1-p_2.
\end{array}\end{eqnarray}
So the above equations (\ref{EqDSupp2}) are the same forms with Eqs.(\ref{EqTdDM}) for the continuous Hilbert space.

\subsection{Fidelities}
The fidelity of the measured overall wave function is defined as below:
\begin{eqnarray}
F_\text{W}=\frac{|\int{\psi(t)\psi_{in}^{*}(t)dt}|}{\sqrt{\int{|\psi(t)|^2dt} \int{|\psi_{in}(t)|^2dt} }},
\end{eqnarray}
where $\psi(t)$  ($\psi_{in}(t)$) is the measured (prepared) wavefunction.

Similarly, the fidelity of the phase $\varphi (t)$ is defined by
\begin{eqnarray}
F_\text{P}=\frac{\int{\varphi(t)\varphi_{in}(t)dt}}{\sqrt{\int{\varphi(t)^2dt} \int{\varphi_{in}(t)^2dt} }},
\end{eqnarray}
where $\varphi(t)$ ($\varphi_{in}(t)$) is the phase of the measured (prepared) wavefunction.

The fidelity of the amplitude $|\psi(t)|$  is defined by
\begin{eqnarray}
F_\text{A}=\frac{\int{|\psi(t)||\psi_{in}(t)|dt}}{\sqrt{\int{|\psi(t)|^2dt} \int{|\psi_{in}(t)|^2dt} }},
\end{eqnarray}
where $|\psi(t)|$($|\psi_{in}(t)|$) is the amplitude of the measured (prepared) wavefunction.

\section{Acknowledgements}
This work was supported by the National Key Research and Development Program of China (Grants No. 2016YFA0302800 and No. 2016YFA0301803), the National Natural Science Foundation of  China (Grants No. 61378012, No. 91636218, No. 11822403, No. 11804104, No. 11804105, No. 61875060 and No. U1801661), the Natural Science Foundation of Guangdong Province (Grant No.2015TQ01X715, No. 2014A030306012, No.2018A030313342 and No. 2018A0303130066), the Key  Project of Science  and  Technology of Guangzhou (Grant No. 201804020055). S. D. acknowledges the support from Hong Kong Research Grants Council (Project No. 16303417) and William Mong Institute of Nano Science and Technology (Project No. WMINST19SC05)
\end{document}